\documentclass[a4paper,11pt]{article}
\pdfoutput=1 

\usepackage{jcappub} 

\usepackage[T1]{fontenc} 

\title{Gravitational waves in a cyclic Universe: resilience through cycles and vacuum state}

\author[a]{Mariaveronica De Angelis,}
\author[a]{Adam Smith,}
\author[a]{William Giar\`e,}
\author[a]{Carsten van de Bruck}

\affiliation[a]{School of Mathematics and Statistics, The University of Sheffield, Hounsfield Road, S3 7RH Sheffield, United Kingdom}

\emailAdd{mdeangelis1@sheffield.ac.uk}
\emailAdd{asmith69@sheffield.ac.uk}
\emailAdd{w.giare@sheffield.ac.uk}
\emailAdd{c.vandebruck@sheffield.ac.uk}

\abstract{
We present a generalised calculation for the spectrum of primordial tensor perturbations in a cyclic Universe, making no assumptions about the vacuum state of the theory and accounting for the contribution of tensor modes produced in the dark energy phase of the previous cycle. We show that these modes have minimal impact on the spectrum observed in the current cycle, except for corrections on scales as large as the comoving Hubble radius today. These corrections are due to sub-horizon modes produced towards the end of the dark energy phase, persisting into the \textit{ekpyrotic} phase of the next cycle as additional quanta. In relation to the vacuum state, we argue that non-Bunch-Davies quanta can easily overwhelm the energy density driving the dark energy phase, potentially compromising the model. Therefore, avoiding backreaction effects sets restrictive constraints on deviations away from the Bunch-Davies vacuum during this phase, limiting the overall freedom to consider alternative vacua in the cyclic Universe.}

\begin{document}
\maketitle
\flushbottom

\section{Introduction}\label{intro}

The most compelling observational evidence supporting cosmological inflation~\cite{Guth:1980zm,Linde:1981mu,Albrecht:1982wi,Vilenkin:1983xq} as the leading theory of the early Universe is currently provided by the Planck satellite measurement of the spectral index of scalar perturbations, $n_s = 0.9649\pm0.0042$~\cite{Planck:2018jri}. In the simplest single-field slow-roll inflationary models, the spectrum of scalar modes is expected to be almost but not exactly flat~\cite{Mukhanov:1981xt, Bardeen:1983qw,Hawking:1982cz,Guth:1982ec}, with deviations from flatness are quantified in terms of how much $n_s$ deviates from 1~\cite{Lidsey:1995np,Lyth:1998xn,Baumann:2009ds,Martin:2013tda}. As a result, the Planck data seem to be in excellent agreement with the theoretical predictions of inflationary models~\cite{Planck:2018jri,Planck:2018vyg}, ruling out a Harrison-Zeldovich scale-invariant spectrum~\cite{Harrison:1969fb,Zeldovich:1972zz,Peebles:1970ag} (corresponding to $n_s=1$) at a statistical level exceeding $8.5$ standard deviations and lending weight to the inflationary paradigm.

That being said, with no aim to downplay the significance of this result or its interpretation, it is crucial to emphasise that, on its own, it does not provide conclusive evidence for cosmological inflation. Even hinging on a certain level of optimism and setting aside the uncertainty surrounding constraints on $n_s$ from CMB experiments other than Planck\footnote{Over the years, constraints on the spectral index have been released by a multitude of Planck-independent CMB experiments such as WMAP~\cite{WMAP:2012fli,WMAP:2012nax}, the Atacama Cosmology Telescope (ACT)~\cite{ACT:2020frw, ACT:2020gnv}, and the South Pole Telescope (SPT)~\cite{SPT-3G:2014dbx, SPT-3G:2021eoc}. When considering these data at face value, Planck is currently the only experiment excluding $n_s=1$ at a statistical significance much larger than $3\sigma$. Conversely, ACT shows a preference for $n_s=1$~\cite{ACT:2020gnv,Giare:2022rvg}. Different combinations of these data overall support the result $n_s\ne1$, although sometimes they lead to discordant results in terms of the other inflationary parameters or the preferred inflationary models~\cite{Forconi:2021que,Giare:2023wzl}.} -- or the potential implications arising from the well-known tensions~\cite{Bernal:2016gxb,Verde:2019ivm,DiValentino:2020zio,DiValentino:2021izs,Abdalla:2022yfr} characterising the recent debate\footnote{For studies suggesting potential implications of cosmological tensions for inflation, see, e.g., Refs.~\cite{DiValentino:2018zjj,Ye:2021nej,Ye:2022efx,Jiang:2022uyg,Jiang:2022qlj,Takahashi:2021bti,Lin:2022gbl,Hazra:2022rdl,Braglia:2021sun,Keeley:2020rmo,Jiang:2023bsz}}
--  alternative theoretical mechanisms have been put forth, yielding an almost scale-invariant spectrum of primordial density fluctuations without invoking inflation.

An illustrative example of such mechanisms is the cyclic Universe~\cite{Steinhardt:2001st,Steinhardt:2002ih,Steinhardt:2002kw,Khoury:2003rt,Turok:2004yx,Khoury:2004xi,Lehners:2008vx} that, in contrast to the conventional cosmological framework, suggests a periodic history for the Cosmos. The model has been extensively studied and discussed in relation to a broad range of topics, including quantum gravity, modified gravity, gravitational waves and dark energy, see e.g., Refs.~\cite{Boyle:2003km,Ashtekar:2003hd,Bojowald:2004kt,Xiong:2007cn,Frampton:2007cv,Narlikar:2007hip,Baum:2007de,Biswas:2008ti,Cailleteau:2009fv,Brandenberger:2009ic,El-Nabulsi:2011xss,Cai:2012ag,Cai:2010zma,Nojiri:2011kd,Chang:2012yk,Ivanov:2012hq,Saaidi:2012qp,Bars:2013vba,Tavakoli:2014mra,Oriti:2016qtz,deCesare:2016rsf,Pavlovic:2017umo,Saridakis:2018fth,Das:2018bzx,Ijjas:2018bko,Ahmed:2019bff,Li:2019laq,Scherrer:2019dkc,Barca:2019ane,Ijjas:2021zwv,Gorkavyi:2021tbw,Martin-Benito:2021szh,Calcinari:2022iss,Giovannetti:2022qje,Giovannetti:2023psb} or Refs.~\cite{Battefeld:2014uga,Brandenberger:2016vhg} for reviews. In broad terms, each cycle comprises a phase recasting the standard Hot Big Bang theory (during which large-scale structures take shape), followed by a phase of slow, accelerated expansion mirroring the present-day observational evidence for a Dark Energy dominated dynamics. In the cyclic Universe, this latter stage also serves to dilute inhomogeneities and flatten the spatial geometry. Subsequently, a contraction phase ensues, generating nearly scale-invariant density perturbations. Finally, the cycle concludes with a big-crunch/big-bang transition, during which matter and radiation are generated, setting the stage for the next cycle. 

Notice that both inflation and the cyclic Universe provide physical mechanisms to produce an almost scale-invariant spectrum of density perturbations~\cite{Khoury:2001zk,Lehners:2007ac,Buchbinder:2007tw}. In addition, they can both explain observational facts such as the homogeneity in the cosmic microwave background (CMB) radiation~\cite{Lehners:2013cka} and the fact that the present-day spatial geometry of the Universe appears to be flat, or at the very least nearly flat\footnote{For recent discussions surrounding the spatial geometry of the Universe, see, e.g.,~\cite{Park:2017xbl,Handley:2019tkm,DiValentino:2019qzk,Efstathiou:2020wem,DiValentino:2020hov,Benisty:2020otr,Vagnozzi:2020rcz,Vagnozzi:2020dfn,DiValentino:2020kpf,Yang:2021hxg,Cao:2021ldv,Dhawan:2021mel,Dinda:2021ffa,Gonzalez:2021ojp,Akarsu:2021max,Cao:2022ugh,Glanville:2022xes,Bel:2022iuf,Yang:2022kho,Stevens:2022evv,Favale:2023lnp}}. Therefore, at first glance, one might wonder how to distinguish between the two models. Focusing solely on scalar modes, this is a challenging knot to unravel~\cite{Khoury:2003vb,Gratton:2003pe}. However, the two scenarios yield significantly distinct predictions for the stochastic background of gravitational waves~\cite{Boyle:2003km}. Similar to scalar modes, inflation predicts a nearly scale-invariant (red-tilted) spectrum of tensor modes~\cite{Baumann:2009ds,Martin:2013tda,Caprini:2018mtu}. Conversely, in the cyclic Universe, the tensor spectrum is typically blue-tilted, and its amplitude is many orders of magnitude lower than that predicted by inflation, remaining well below any observable threshold achievable in the near future. Consequently, any measurement of primordial gravitational waves (e.g., through the effects left in the CMB B-mode polarisation at large angular scales) would offer conclusive evidence for inflation, discounting the cyclic model. 

Despite this fact being acknowledged as a strength for inflation and perhaps a limitation in predictive capacity for the cyclic model, it is worth emphasising a few caveats surrounding this conclusion. Firstly, despite the best efforts, the detection of primordial tensor modes remains elusive at present~\cite{BICEP:2021xfz}, making it impossible to discriminate between the two scenarios. Therefore, the cyclic Universe remains an alternative worth considering. Secondly, the inflationary predictions concerning the amplitude and tilt of the tensor spectrum depend significantly on the specific model. While well-known consistency relations among inflationary parameters can be derived within single-field slow-roll inflation minimally coupled to gravity~\cite{Baumann:2009ds,Martin:2013tda}, these relations can be violated by various physical mechanisms. A long yet not exhaustive list of possibilities include considering modified gravity~\cite{Kobayashi:2010cm,Kawasaki:2013xsa,Nozari:2016jmn,Giare:2020vss}, multi-field inflation~\cite{Kaiser:2013sna,Price:2014ufa,Achucarro:2010da,DeAngelis:2023fdu,Giare:2023kiv}, additional (spectator) rolling axion fields~\cite{Mukohyama:2014gba,Namba:2015gja,Peloso:2016gqs,Ozsoy:2020ccy}, couplings to axion-gauge or spin-2 fields~\cite{Dimastrogiovanni:2016fuu,Iacconi:2019vgc}, breaking spatial and/or temporal diffeomorphism invariance~\cite{Endlich:2012pz,Cannone:2014uqa,Graef:2015ova,Ricciardone:2016lym}, higher curvature corrections to the effective gravitational action~\cite{Baumann:2015xxa,Giare:2020plo}, higher order operators in effective field theory~\cite{Capurri:2020qgz,Giare:2022wxq}, violations of the null energy condition~\cite{Cai:2022lec,Ye:2023tpz}, alternative vacuum state/initial conditions~\cite{Ashoorioon:2013eia,Ashoorioon:2014nta,Choudhury:2023kam}, sound speed resonances~\cite{Cai:2020ovp}, inflation in an Universe filled with an elastic medium~\cite{Gruzinov:2004ty}, and possible effects/models inspired by quantum gravity~\cite{Ashoorioon:2005ep,Brandenberger:2006xi,Brandenberger:2014faa,Baumgart:2021ptt}. Many of these more elaborated scenarios yield completely different predictions, often resulting in a blue-tilted spectrum and possibly leaving signatures in different cosmological and astrophysical observables~\cite{Stewart:2007fu,Cai:2014uka,Wang:2014kqa,Kuroyanagi:2014nba,Kuroyanagi:2020sfw,Giare:2020vhn,Vagnozzi:2020gtf,Vagnozzi:2023lwo,Jiang:2023gfe,Oikonomou:2024aww}. Furthermore, models with an arbitrarily small tensor amplitude can always be constructed (see, e.g., Ref~\cite{Stein:2022cpk}), making it virtually impossible to rule out inflation based solely on a lack of detection of primordial gravitational waves. This is a critique frequently raised against inflationary cosmology as it questions its actual predictive capability.

As concerns the cyclic Universe, since any difference with respect to inflation in terms of predictions is likely to be confined to the spectrum of tensor modes, it becomes interesting to test whether similar caveats apply or if the model demonstrates greater resilience. 

In light of this, we review the production of primordial gravitational waves in a cyclic Universe, identifying (and eventually clarifying) some conceptual aspects related to its concrete predictivity. Specifically, prevailing calculations in the existing literature conventionally establish initial conditions for primordial scalar and tensor modes during the \textit{ekpyrotic} contracting phase~\cite{Boyle:2003km}. While for scalar perturbations the implications of setting the initial conditions in different phases have been examined~\cite{Erickson:2006wc}, the calculation of the tensor spectrum has always been performed starting in the \textit{ekpyrotic} phase, assuming a Bunch-Davies (BD) vacuum state and neglecting any potential contributions arising from tensor modes originating during the dark energy phase of the previous cycle. This leads us to question whether they exert any influence on the spectrum observed in the current cycle. Taking a broader perspective, one may wonder whether the predictions concerning tensor modes remain resilient throughout the diverse cycles of the model itself. Yet another aspect that is imperative to clarify is to what extent the predictions depend on the choice of the vacuum state, addressing the crucial question of what freedom exists in the cyclic Universe regarding the choice of the vacuum state and whether substantial alterations can arise in the tensor spectrum by assuming different vacua, akin to what is found in inflationary cosmology.

To address these points, we present a general model for the evolution of gravitational waves produced in a cyclic Universe, making no assumptions about the initial vacuum state and starting the calculation from the dark energy phase of the previous cycle. We find that the additional tensor models originated in the previous cycle have minimal impact on the tensor spectrum observed in the current cycle, except for corrections on scales as large as the comoving Hubble radius today that are due to sub-horizon modes produced towards the end of the previous dark energy phase. Most importantly, we find that non-BD quanta in the dark energy phase can easily overwhelm the energy density associated with the modulus field, potentially spoiling the model. Avoiding these backreaction effects sets restrictive constraints on deviations away from the BD vacuum during the dark energy phase, thereby limiting the overall freedom to consider alternative vacua in the cyclic Universe.

The paper is organised as follows. In \autoref{construction of the model}, we introduce the cyclic Universe model and review its background dynamics. In \autoref{primordial tensor spectrum}, we consider the evolution of gravitational waves in such a Universe, starting from the previous cycle's dark energy phase and deriving the evolution in full generality. In \autoref{sec:discussion}, we discuss the implications for the model's predictions, deriving constraints on the choice of the vacuum state and analysing the contribution coming from tensor modes originated in the previous cycle. Finally, in \autoref{sec:conc}, we derive our main conclusions.

\section{Cyclic model and Background Dynamics} \label{construction of the model}
We consider a simple scalar field setup in which the dynamics of the cyclic model in the Einstein frame are well described by the 4$D$ effective Lagrangian~\cite{Erickson:2006wc}
\begin{equation}\label{lagrangian}
    \mathcal{L}=\sqrt{-g}\left(\frac{M_{\rm Pl}^2}{2}R-\frac{1}{2}\partial_\mu\phi\partial^\mu\phi-V(\phi)\right),
\end{equation}
where $g$ is the determinant of the metric $g_{\mu \nu}$, $R$ is the Ricci scalar and we adopt units where $c=1$. The scalar field $\phi$ is a modulus field, driving the dark energy dominated phase and the subsequent \textit{ekpyrotic} and contracting kinetic phases, which we discuss below. Assuming a spatially flat FRLW background, the scalar field satisfies the usual equation of motion
\begin{equation}
    \ddot{\phi}+3H\dot{\phi}+V_{,\phi}=0,
    \label{KG}
\end{equation}
where dots denote derivatives with respect to cosmological time $t$. On the other hand, ignoring any coupling between the scalar field and the other standard model species and neglecting any additional contributions from the latter to the total Universe energy density, the evolution of the scale factor is governed by the Friedmann equation that, in terms of the Hubble parameter $H=\dot{a}/a$, reads
\begin{equation}
    H^2=\frac{1}{3 M_{\rm Pl}^2}\biggl(\frac{1}{2}\dot{\phi}^2 + V(\phi)\biggl).
    \label{H2}
\end{equation}
In what follows, to efficiently describe the dynamics of the cyclic Universe we focus on a phenomenological potential of the form
\begin{equation}\label{potnetial} 
    V=V_0\left(1-e^{-c\phi/M_{\rm Pl}}\right)Y(\phi),
\end{equation}
where $V_0$ is of the same order of the vacuum energy observed in today’s Universe, $c$ is a positive constant value and $Y(\phi)$ is a step function. Notice that our choice concerning the specific potential employed in the work is, in part, motivated by the fact that the exponential form is convenient for analysis, and in part from the fact that the same potential has been widely adopted in similar studies, allowing a direct comparison between our findings and other results documented in the existing literature. However, it is important to emphasise that cyclic models can emerge from a broad spectrum of different potentials that should ultimately emerge from the higher-dimensional theory. Without loss of generality, the only constraint comes from requiring an acceptable spectrum of scalar perturbations that implies considering a steep, strongly negatively curved region across observational ranges of the field to reproduce. 

Having that said, the potential~\eqref{potnetial} serves multiple purposes, including describing dark energy responsible for cosmic acceleration observed today. More importantly, it plays a crucial role in transitioning the Universe from accelerated expansion to contraction. This is achieved by rolling from positive to negative values of the potential until reaching a time where $H^2=0$ and consequently triggering a phase characterised by an equation of state $\omega \gg 1$. For instance, by solving Eq.~\eqref{H2} it can be shown that when the negative potential dominates $V\simeq -V_0e^{-c\phi}$ (\textit{i.e.} \textit{ekpyrotic} phase), the scale factor behaves as~\cite{Erickson:2006wc}
\begin{equation}
    a(t)\sim (-t)^{\tilde{\alpha}},
\end{equation}
where $t$ has negative values and $\tilde{\alpha}=2/c^2$. At this point it is also convenient to introduce the conformal time $d\tau=dt/a(t),$ which we will frequently use later. In terms of the conformal time, the scale factor during the \textit{ekpyrotic} phase evolves as:
\begin{equation}
    a(\tau)\sim \biggl[(-1)^{\tilde{\alpha}}(\tau-\tilde{\alpha} \tau)^{\tilde{\alpha}/(1-\tilde{\alpha})}\biggl]_{\tau_i}^{\tau_{f}},
    \label{scalefactor}
\end{equation}
underscoring that the Universe is gradually contracting while the scalar field slowly descends along its sharply decreasing negative potential, to produce an acceptable spectrum of cosmological scalar perturbations.

In the literature, the \textit{ekpyrotic} phase is typically assumed as the starting point of the cycle where initial conditions of primordial perturbations are imposed, and the calculations of the relative scalar and tensor spectra begin. However, in this study, we want to extend the model to include the contribution of tensor perturbations produced during the dark energy phase of the previous cycle to investigate whether they could have any impact on the spectrum we observe in the current cycle and eventually clarify why (not). To do so, the overall strategy will be to start the calculation in the dark energy phase of the previous cycle (making no assumptions on the vacuum state) and evolve the system through four regimes. For this reason, before dealing with the explicit calculation of the tensor spectrum, given that in our case we consider one more phase than in previous studies, it is useful to dedicate the following two subsections to reviewing the background dynamics of the model. In the same spirit of the discussion outlined so far, we start from the dark energy phase of the previous cycle and evolve the scale factor, ensuring its continuity across the boundaries of each phase. Additionally, we derive constraints on the model's parameters based on minimum theoretical requirements, such as the continuity of $H(t)$ and the consistency of the theory across cycles.

\subsection{Evolution and continuity of the scale factor across stages}

\subsubsection{Dark Energy phase}
We start from the dark energy phase, in which the expansion rate $H$ is roughly constant and the scale factor $a(t)$ behaves as
\begin{equation}
    a(t)=a(t_{\rm{tr}})e^{H\left(t-t_{\rm{tr}}\right)} \;\;\;\;t<t_{\rm{tr}},
    \label{adarkenergy}
\end{equation}
with $t_{tr}$ transition time between dark energy and \textit{ekpyrotic} phase. The equation above can be translated in terms of conformal time as
\begin{equation}
    a(\tau) = \frac{1}{H(B-\tau)},
    \label{adarkenergyconf}
\end{equation}
where, for the continuity across $\tau = \tau_{\rm{tr}}$, $B$ is fixed to\footnote{The value of $B$ is achieved by considering $H_0, a(\tau_{\rm{r}})$ and $\tau_{\rm{r}}$ given in the next sections.}
\begin{equation}\label{B_const}
    B = \frac{1}{a(\tau_{\rm{tr}})H}+\tau_{\rm{tr}}.
\end{equation}
Additionally, by means of Eq.~\eqref{scalefactor}, we can infer 
\begin{equation}
    \frac{a(\tau_{\rm{tr}})}{a(\tau_{\rm{end}})} = \left(\frac{\tau_{\rm{tr}}- \tau_{\rm{ek}}}{\tau_{\rm{end}}-\tau_{\rm{ek}}}\right)^\alpha,
\end{equation}
where $\alpha\equiv \tilde{\alpha}/(1-\tilde{\alpha})$ and $\tau_{\rm{ek}}\equiv (1-2\alpha)\tau_{\rm{end}}$ is the conformal time corresponding to when the potential diverges to minus infinity.

\subsubsection{Ekpyrotic phase}

As a next step, we transition to the \textit{ekpyrotic} phase. In this phase the potential becomes negative and the Einstein frame expansion forces the scale factor to contract
\begin{equation}\label{ekpyrotic scale factor}
    \frac{a(\tau)}{a(\tau_{end})}=\left(\frac{\tau-\tau_{ek}}{\tau_{end}-\tau_{ek}}\right)^\alpha\;\;\;\;\; \tau_{tr}<\tau<\tau_{end}.
\end{equation}
We note again that, being $\alpha\ll 1$, the contraction is very slow.

\subsubsection{Contracting kinetic phase}

Once $\tau > \tau_{\rm{end}}$ we enter the region where the effects of the potential are negligible, namely $\phi<\phi_{\rm{end}}$. During this period -- known as contracting kinetic phase -- we have:
\begin{equation}
    \frac{a(\tau)}{a(\tau_{\rm{r}})}=\left(\frac{-\tau}{(1+\chi)\tau_{\rm{r}}}\right)^{\frac{1}{2}}\;\;\;\;\;\tau_{\rm{end}}<\tau<0,
    \label{akineticcontraction}
\end{equation}
where $\chi$ is a small positive constant that measures the amount of radiation created at the bounce ($\tau=0$). 

\subsubsection{Expanding kinetic phase}

Finally, for the last phase of the cycle, the so-called expanding kinetic phase, we get
\begin{equation}
    a(\tau)=\left(\frac{\tau}{\tau_{\rm{r}}}\right)^{\frac{1}{2}},\;\;\;\;\;0<\tau<\tau_{\rm{r}},
    \label{aexpanding}
\end{equation}
Notice that, for convenience, we work in a coordinate frame where the scale factor is set to $a(\tau_{\rm{r}}) = 1$ for the time $t_{\rm{r}}$ corresponding to the beginning of the radiation-dominated era. The corresponding conformal time $\tau_{\rm{r}}= (2H_{\rm{r}})^{-1}$ is constrained by the radiation temperature $T_{\rm{r}}$, being $H_{\rm{r}}\propto T_{\rm{r}}^2/M_{\rm{Pl}}$. Following Ref.~\cite{Khoury:2003rt}, we choose a quite conservative value  $T_{\rm{r}} \sim 10^7$ GeV, akin to that obtained in the more familiar standard cosmology at the end of the reheating phase following inflation. This choice also ensures that we can safely recover predictions of primordial Big Bang Nucleosynthesis (BBN).

\subsection{Parameter constraints}

\subsubsection{Continuity of the Hubble parameter}

In order to constrain the length of the \textit{ekpyrotic} contracting phase, we require that the Hubble parameter returns to its original value after every cycle. Following Ref.~\cite{Erickson:2006wc}, this implies that
\begin{equation}
H_{\rm{end}}/H_{\rm{0}}\approx \sqrt{\frac{-V_{\rm{end}}}{V_0}}
\end{equation}
where $V_{\rm{end}}$ is the depth of the potential well and $V_0$ is the height of the potential plateau. The spectral range of perturbations produced when the field rolls from $V\approx 0$ to $V\approx -V_{\rm{end}}$, satisfies
\begin{equation}
    \frac{k_{\rm{max}}}{k_{\rm{min}}}\approx\sqrt{\frac{-V_{\rm{end}}}{V_0}},
    \label{Hratio}
\end{equation}
and it needs to span at least $N=60$ e-folds for the \textit{ekpyrotic} phase to produce a scale-invariant spectrum over a broad enough range of scales for us to observe today~\cite{Planck:2018jri,Planck:2018vyg}. Hence, the transition time between dark energy and \textit{ekpyrotic} phase is constrained to
\begin{equation}
    \frac{H_{\rm{tr}}}{H_{\rm{end}}}=2\, \alpha\, \tau_{\rm{end}} \, \frac{a_{\rm{end}}\, a_{\rm{tr}}}{{(\tau_{\rm{tr}}-\tau_{\rm{ek}})}}<e^{-60}.
    \label{Htrans/Hend}
\end{equation}
Moreover, as $a(\tau)\approx \rm{const}$ during the \textit{ekpyrotic} contracting phase \cite{Erickson:2006wc}, from Eq.~\eqref{Htrans/Hend} it follows that
\begin{equation}
    \left|\tau_{\rm{tr}}-\tau_{\rm{ek}}\right|>2\, \alpha\, \tau_{\rm{end}}\, a_{\rm{end}}^2\, e^{60}.
\end{equation}

\subsubsection{Cycling constraint}
To place constraints on the duration of the kinetic evolution phases, we require that they must last enough time for the scalar field to have started at the potential minimum (\textit{i.e.} $\phi=\phi_{\rm{end}}$), moved off to the Bounce (\textit{i.e.} $\phi\to-\infty$), and returned all the way back past $\phi_{\rm{end}}$ and made it up to the potential plateau to begin a radiation-dominated Universe.

Barring some brief $\omega\gg 1$ period (which divides the expanding kinetic phase into two parts) as the field moves back up past $\phi_{\rm{end}}$ to the potential plateau, from Eq.~\eqref{KG} and Eq.~\eqref{potnetial} we find
\begin{equation}
    \phi-\phi_{end}=c_1 \ln\left(\frac{t}{t_{end}}\right),
\end{equation}
where the factor $c_1^2=2/3$ comes from Eq.~\eqref{H2} during kinetic domination, $t_{\rm{end}}$ is the time taken to reach $\phi_{\rm{end}}$  starting at $\phi\to-\infty$. 

Notice that in the region where $\omega\gg 1$, contributions from $Y(\phi)$ becomes relevant and $V\approx V_0\left(1-e^{-c\phi}\right)$. Therefore, in this case, the time $t_{\rm{r}}$ required to climb the potential well and reach the plateau at $V\approx V_0$ can be bounded to 
\begin{equation}
    \frac{t_{\rm{r}}}{t_{\rm{end}}}>\left(\frac{V_{\rm{end}}}{V_0}\right)^{\sqrt{\frac{3}{2 c^2}}}.
    \label{eq:tr_tend}
\end{equation}
Since the time taken for the field to cross the negative region of the potential before radiation domination begins is given by
\begin{equation}
    \frac{t_{\rm{r}}}{t_{\rm{end}}}\approx \frac{\sqrt{V_{\rm{end}}}}{H_{\rm{r}}},
\end{equation}
from Eq.~\eqref{eq:tr_tend} we can infer an upper limit on the Hubble parameter at $t_r$ which reads  
\begin{equation}
    H_{\rm{r}} \lesssim \frac{\sqrt{V_{\rm{end}}}}{M_{\rm{Pl}}}\left(\frac{V_0}{V_{\rm{end}}}\right)^{\frac{3}{2c^2}}.
    \label{Hr}
\end{equation}    
This upper limit constrains the ratio between $\tau_r$ and $\tau_{\rm end}$ to
\begin{equation}\label{gamma}
    \Gamma = \left|\frac{\tau_{\rm{r}}}{\tau_{\rm{end}}}\right| \gtrsim \left(\frac{V_{\rm{end}}}{V_0}\right)^{\sqrt{\frac{2}{3c^2}}}\simeq 10^8,
\end{equation}
where we used that the vacuum energy density $\rho_{\Lambda} = V_0 \sim 10^{-120}M_{\rm{Pl}}^4$ and $V_{\rm{end}}\sim 10^{-20} M_{\rm{Pl}}^4$.
Following Ref.~\cite{Boyle:2003km}, throughout this paper, we always consider the dimensionless parameter $\Gamma \sim 10^8$.\\
\\
\noindent In conclusion, the final constraints we derive for the cyclic model at the background level (and that are important to bear in mind for the following discussion on tensor perturbations) are:
\begin{align}
    \left|\frac{\tau_{\rm{r}}}{\tau_{\rm{end}}}\right|&\gtrsim 10^8, & (\tau_{\rm{tr}} - \tau_{\rm{ek}})&\gtrsim 10^9, & \tau_{\rm{r}} = \frac{1}{2H_{\rm{r}}}.
\end{align}

\section{General primordial tensor spectrum}
\label{primordial tensor spectrum}
Considering a spatially flat FLRW metric, in the synchronous gauge the perturbed line element reads:
\begin{equation}
d s^2=a^2(\tau)\left[d \tau^2-\left(\delta_{i j}+h_{i j}\right) d x^i d x^j\right].
\end{equation}
Tensor modes (i.e., metric perturbations) are described in terms of the transverse and traceless part of the symmetric 3×3 matrix $h_{ij}$. To characterise the contribution of each wavenumber $k$ to $h_{ij}(t,\mathbf{x})$, we consider a Fourier representation $\tilde{h}_{ij}(t,\mathbf{k})$. Moving to the Fourier space, focusing on one particular polarisation state, and assuming isotropy, the gravitational wave field $h_{k}$ satisfies the following equation:
\begin{equation}
    h_{k}'' + 2\frac{a'}{a}h_k'+k^2h_k=0,
\end{equation}
where $(..)'$ denotes the derivative with respect to conformal time $\tau$. However, it is more convenient to use a new variable $f_k(\tau)\equiv a(\tau)h_k(\tau)$ satisfying
\begin{equation}
\label{BesselsEq}
f_k''+\left(k^2+\frac{a''}{a}\right)f_k=0.
\end{equation}
After redefining $f_k=i\sqrt{\tau}\,u_k$, Eq.~\eqref{BesselsEq} assumes the more familiar form of a Bessel equation and, for each phase of the cyclic model, the general solution involves a linear combination of the Hankel functions of the first and second kind, that we denote as $H^{(1,2)}$. 

In this section, we present a generalised calculation for the primordial tensor spectrum in the cyclic Universe. In \autoref{sec.general_solution}, we derive the general solution of Eq.~\eqref{BesselsEq} making no assumptions about the vacuum state of the theory and considering the contribution of tensor modes produced in the dark energy phase of the previous cycle rather than starting directly from the \textit{ekpyrotic} phase of the present cycle. In \autoref{sec.Matching_phases}, we require internal consistency in the evolution of the tensor mode amplitudes throughout the four phases of the model, ensuring that both $h_k(\tau)$ and $h_k'(\tau)$ remain continuous functions and matching the general solutions across the different phases. Finally, in \autoref{sec.spectrum_today}, we derive the expression of the primordial tensor spectrum and briefly discuss its strain today.

\subsection{General Solutions in the different Phases}

\label{sec.general_solution}

\subsubsection{General solution in the Dark Energy phase}
During the dark energy phase, $a(\tau)$ is given by Eq.~\eqref{adarkenergyconf}, and the general solution of Eq.~\eqref{BesselsEq} reads 
\begin{equation}\label{DE solution}
    f_k(\eta)=\sqrt{-k\eta}\left(D_1(k)H^{(1)}_\frac{3}{2}(-k\eta)+D_2(k)H^{(2)}_{\frac{3}{2}}(-k\eta)\right)\;\;\;\tau<\tau_{\rm{tr}},
\end{equation}
where $H_n^{(1)}$ and $H_n^{(2)}$ denote the Hankel functions of first and second kind respectively, $\eta=\tau-B$, where $B$ is given by Eq.~\eqref{B_const} and, for each wave-number $k$, $D_{1,2}(k)$ are arbitrary constants. 

\subsubsection{General solution in the \textit{ekpyrotic} contraction}
In the stage of the \textit{ekpyrotic} contraction, $a(\tau)$ is given by Eq.~\eqref{ekpyrotic scale factor} and the general solution reads
\begin{equation}\label{ekpyrotic solution}
f_k(\tau)=\sqrt{y}\left(A_1(k)H^{(1)}_n(y)+A_2(k)H^{(2)}_{n}(y)\right)\;\;\;\tau_{\rm{tr}}<\tau<\tau_{\rm{end}},
\end{equation}
where $y\equiv-k(\tau-\tau_{\rm{ek}})$. 

\subsubsection{General solution in the kinetic contracting phase}
Moving to the kinetic contracting phase and making use of Eq.~\eqref{akineticcontraction}, we achieve
\begin{equation}
    f_k(\tau)=\sqrt{-k\tau}\left(B_1(k)H^{(1)}_0(-k\tau)+B_2(k)H^{(2)}_{0}(-k\tau)\right)\;\;\;\tau_{\rm{end}}<\tau<0,
\end{equation}

\subsubsection{General solution in the kinetic expanding phase}
Considering Eq.~\eqref{aexpanding} in the kinetic expanding phase the genral solution is
\begin{equation}
    f_k(\tau)=\sqrt{k\tau}\left(C_1(k)H^{(1)}_0(k\tau)+C_2(k)H^{(2)}_{0}(k\tau)\right)\;\;\;0<\tau<\tau_{\rm{r}}.
\end{equation}

\subsection{Matching phases}
\label{sec.Matching_phases}
In what follows we obtain expressions for the coefficients $D_{1,2}(k),A_{1,2}(k),B_{1,2}(k)$ and $C_{1,2}(k)$ by matching $h_k(\tau)$ and $h_k(\tau)'$ at the boundaries of each phase. \\

\subsubsection{Dark energy - \textit{ekpyrotic}}
At the boundary between dark energy and \textit{ekpyrotic} stage, we require continuity of $f_k(\tau)$ namely
\begin{equation}\label{dark energy-ekpyrotic matching}
    \sqrt{x_{\rm{tr}}}\left(D_1(k)H^{(1)}_\frac{3}{2}(x_{\rm{tr}})+D_2(k)H^{(2)}_{\frac{3}{2}}(x_{\rm{tr}})\right)= \sqrt{y_{\rm{tr}}}\left(A_1(k)H^{(1)}_n(y_{\rm{tr}})+A_2(k)H^{(2)}_{n}(y_{\rm{tr}})\right),
\end{equation}
where $x_{\rm{tr}}=-k\eta(\tau_{\rm{tr}})$, and continuity of $f'(\tau)$ which we write in the matrix form $\underline{\underline{\textbf{y}}}_1\textbf{A}=\underline{\underline{\textbf{x}}}_1\textbf{D}$ where
    \\

\begin{align}\label{transition matching}
 \underline{\underline{\textbf{y}}}_1&=\left(
 \renewcommand*{\arraystretch}{1.5}
    \begin{matrix}
        \sqrt{y_{\rm{tr}}}H^{(1)}_n(y_{\rm{tr}}) & 
        \sqrt{y_{\rm{tr}}}H^{(2)}_n(y_{\rm{tr}})\\[5pt]
        \sqrt{y_{\rm{tr}}}H^{(1)}_{n-1}(y_{\rm{tr}})-\frac{(n-\frac{1}{2})}{\sqrt{y_{\rm{tr}}}}H^{(1)}_n(y_{\rm{tr}})
        & \qquad
        \sqrt{y_{\rm{tr}}}H^{(2)}_{n-1}(y_{\rm{tr}})-\frac{(n-\frac{1}{2})}{\sqrt{y_{\rm{tr}}}}H^{(2)}_n(y_{\rm{tr}})
    \end{matrix}
    \right),\nonumber\\[10pt]
    \underline{\underline{\textbf{x}}}_1&=\left(
    \begin{matrix}
        \sqrt{x_{\rm{tr}}}H^{(1)}_\frac{3}{2}(x_{\rm{tr}}) & \sqrt{x_{\rm{tr}}}H^{(2)}_\frac{3}{2}(x_{\rm{tr}})\\[5pt]
        \sqrt{x_{\rm{tr}}}H^{(1)}_\frac{1}{2}(x_{\rm{tr}})-\frac{1}{\sqrt{x_{\rm{tr}}}}H^{(1)}_\frac{3}{2}(x_{\rm{tr}})
        & \qquad
        \sqrt{x_{\rm{tr}}}H^{(2)}_\frac{1}{2}(x_{\rm{tr}})-\frac{1}{\sqrt{x_{\rm{tr}}}}H^{(2)}_\frac{3}{2}(x_{\rm{tr}})
    \end{matrix}
    \right),
\end{align}
and
\begin{equation}
    \textbf{A} = \left(\begin{matrix}
        A_1\\A_2
    \end{matrix}\right),\;\;\;\;\;\;
    \textbf{D} = \left(\begin{matrix}
        D_1\\D_2
    \end{matrix}\right),\;\;\;\;\;\; \text{with} \;\; x_{\rm{tr}}\equiv-k\tau_{\rm{tr}}.
\end{equation}

\subsubsection{\textit{Ekpyrotic} - kinetic contraction}

Similarly here, at $\tau = \tau_{\rm{end}}$ namely the end of the \textit{ekpyrotic} phase, we require $\underline{\underline{\textbf{y}}}_2\textbf{B}=\underline{\underline{\textbf{x}}}_2\textbf{A}$ where 
\begin{align}
 \underline{\underline{\textbf{y}}}_2&=\left(
    \begin{matrix}
        H^{(1)}_0(x_{\rm{e}}) & H^{(2)}_0(x_{\rm{e}})\\[5pt]
        \sqrt{x_{\rm{e}}}H^{(1)}_{-1}(x_{\rm{e}})+\frac{H^{(1)}_0(x_{\rm{e}})}{2\sqrt{x_{\rm{e}}}}
        &\qquad
        \sqrt{x_{\rm{e}}}H^{(2)}_{-1}(x_{\rm{e}})+\frac{H^{(2)}_0(x_{\rm{e}})}{2\sqrt{x_{\rm{e}}}}
    \end{matrix}
    \right),\nonumber \\[10pt]
    \underline{\underline{\textbf{x}}}_2&=\left(
    \begin{matrix}
        \sqrt{2\alpha}H^{(1)}_n(2\alpha x_{\rm{e}}) & 
        \sqrt{2\alpha}H^{(2)}_n(2\alpha x_{\rm{e}})\\[5pt]
        \sqrt{2\alpha x_{\rm{e}}}H^{(1)}_{n-1}(2\alpha x_{\rm{e}})-\frac{(n-\frac{1}{2})}{\sqrt{2\alpha x_{\rm{e}}}}H^{(1)}_n(2\alpha x_{\rm{e}})
        & \qquad
        \sqrt{2\alpha x_{\rm{e}}}H^{(2)}_{n-1}(2\alpha x_{\rm{e}})-\frac{(n-\frac{1}{2})}{\sqrt{2\alpha x_{\rm{e}}}}H^{(2)}_n(2\alpha x_{\rm{e}})
    \end{matrix}
    \right),
\end{align}
and 
\begin{equation}
    \textbf{B}=\left(\begin{matrix}
    B_1\\B_2
\end{matrix}\right),\;\;\;\;\;\; \text{with} \;\; x_{\rm{e}}\equiv-k\tau_{\rm{end}}.
\end{equation}

\subsubsection{Kinetic contraction- kinetic expansion}
The matching for these two final stages arises at $\tau = 0$, which is trivial as we have
\begin{equation}
    C_{1,2} = -\sqrt{1+\chi}B_{2,1}.
\end{equation}

\subsection{Strain spectrum today ($\tau=\tau_0$)}
\label{sec.spectrum_today}

The quantity in terms of which we assess the production of primordial gravitational waves in the cyclic Universe is the dimensionless strain spectrum 
\begin{equation}\label{general_strain_expreession}
    \Delta h = k^{\frac{3}{2}}\frac{\left|h_k(\tau)\right|}{\pi}.
\end{equation}
It is convenient to evaluate the dimensionless strain spectrum in the radiation-dominated epoch. Using the general solutions we derived and matched across the different phases of the model, it reads:
\begin{equation}\label{strain_spectrum}
    \Delta h(k, \tau_{r}) = \frac{k^2\sqrt{2\,\tau_{\rm{r}}}}{a(\tau_{\rm{r}})\pi M_{\rm{Pl}}}\left|C_1(k) H_0^{(1)}(x_{\rm{r}}) + C_2(k)H_0^{(2)}(x_{\rm{r}})\right|,
\end{equation}
where $x_{\rm{r}}\equiv k\,\tau_{\rm{r}}$. The strain tensor spectrum today $\Delta h(k, \tau_0)$ can be easily related to the spectrum in the radiation dominated epoch by means of the transfer function formalism as:
\begin{equation}\label{spectrum today}
    \Delta h(k, \tau_0) \equiv \mathcal{T}(k) \Delta h(k, \tau_{\rm{r}}), 
\end{equation}
where $\mathcal{T}(k)$ is the transfer function, responsible for propagating the spectrum forwards to today's observed spectrum. Following Ref.~\cite{Boyle:2003km}, we use a transfer function of the form 
\begin{equation}\label{transfer function}
    \mathcal{T}(k)\approx \left(\frac{k_0}{k}\right)^2\left(1+\frac{k}{k_{\rm{eq}}}+\frac{k^2}{k_{\rm{eq}}k_{\rm{r}}}\right),
\end{equation}
where $k_{\rm{r}} = a_{\rm{r}} H_{\rm{r}} \approx T_r^2/M_{\rm{Pl}}
$
is the wave number of modes crossing the horizon at the start of radiation domination. 

So far, the whole calculation is done by setting $a(\tau_{\rm{r}})=1$. However, for ease of comparison with data and the other results documented in the literature, it is convenient to return to the usual coordinate system where $a_0 = 1$. This can be easily done by means of the well-known inverse scaling between the scale factor and the CMB temperature (\textit{i.e.} $a \propto 1/T$). This relation fixes the ratio $a(\tau_{0}) / a(\tau_{\rm{r}}) \propto T_{\rm{r}}/T_0 \sim 10^{20}$. As a result, we can re-scale the comoving frequency accordingly, which now takes the more familiar value $\hat k_{\rm{r}}=k_{\rm{r}}/a(\tau_{0})=10^{-1}\, \text{Hz}$. From now on, we will name the re-scaled comoving frequency $\hat k_{\rm{r}}$ as $k_{\rm{r}}$. Additionally, always following from the inverse proportionality between the scale factor and the CMB temperature, we expect a spectral range of modes entering between the start of radiation domination and today of the order of
\begin{equation}
    \frac{k_0}{k_{\rm{r}}}\propto \frac{T_0}{T_{\rm{r}}} \sim 6.6 \times 10^{-20}.
\end{equation}
Taking in mind that during matter domination $H\sim a^{-3/2}$, we get
\begin{equation}
    \frac{k_{\rm{eq}}}{k_0}\approx \sqrt{1+z_{\rm{eq}}} \sim 10^2,
\end{equation}
where $z_{\rm{eq}}$ is the redshift at the equivalence.

We can also place a bound on the wave number of modes on the horizon at the previous dark energy-\textit{ekpyrotic} transition, $k_{\rm{tr}}$, which will be useful later. The wavelength of such modes can be estimated using the duration of the \textit{ekpyrotic} phase, which we require to last at least 60 e-folds to effectively homogenise and flatten the Universe for the subsequent cycle. However, a much stronger constraint comes from Eq.~\eqref{Hratio}, as the number of e-folds during the \textit{ekpyrotic} phase is given by $N\sim \ln(H_{\rm{end}}/H_0)\approx \ln(\sqrt{-V_{\rm{end}}/V_0})$. Taking our previous constraints on $V_0$ at today's dark energy density and $V_{\rm{end}}$ at the GUT scale, we arrive at $N\sim 115$. This forces our horizon to shrink by a factor of $\sim 10^{50}$ during the dark \textit{ekpyrotic} phase, and hence we would expect
\begin{equation}
\frac{k_{\rm{end}}}{k_{\rm{tr}}}\approx 10^{50}.
\end{equation}

\section{Resilience through cycles and the vacuum state}
\label{sec:discussion}
The calculation of the spectrum of gravitational waves introduced in the previous section, in addition to considering the contribution of tensor modes produced in the dark energy phase of the previous cycle, is entirely general regarding the vacuum state and applies to any particular choice. This allows us to test several conceptual aspects of the theory (partially highlighted in the introduction), such as its resilience throughout cycles and the implications of the choice of the vacuum states. In this section, we examine both these issues in more detail. In \autoref{Gravitational Waves From A non-Bunch Davies Vacuum}, we focus on the choice of the vacuum state of the theory, deriving novel constraints on the latter based on internal consistency through cycles and epochs. In \autoref{analysis of gravitational waves}, we quantify the extent to which including the contribution from tensor modes originating during the dark energy phase of the previous cycle changes the spectrum of tensor perturbations observed in the current cycle.

\subsection{Gravitational waves from a non-Bunch Davies vacuum}\label{Gravitational Waves From A non-Bunch Davies Vacuum}

As often speculated both in quantum field theory and effective field theory, the choice of the vacuum state represents an important topic of discussion and significance as it plays a crucial role in determining the properties of the theory and the physical predictions it makes. On the one hand, in quantum field theory, the vacuum state is the lowest-energy state of a quantum field, which usually corresponds to the state with no physical particles. On the other hand, effective field theories often deal with specific energy scales or regimes of a more fundamental theory (such as in our effective representation of the cyclic Universe), and the choice of the vacuum state may involve integrating out high-energy degrees of freedom and focusing on the low-energy behavior. This makes the choice of the vacuum state far from trivial when considering the possibility of new physics at sufficiently high energies. 

The selection of the vacuum state has been the subject of intense study and attention in the context of inflationary models where considering more exotic (though physically motivated) vacuum states other than BD) can lead to markedly different predictions for the spectrum of scalar and tensor modes, see \textit{e.g.} ~\cite{Ashoorioon:2013eia,Ashoorioon:2014nta,Choudhury:2023kam} and references therein. Just to mention one example among many, even sticking to the framework of single-field inflation minimally coupled to gravity, considering an exotic vacuum can result in the violation of the usual slow-roll consistency relations allowing for a blue-tilted spectrum of gravitational waves, with an enhanced amplitude on small scales that can eventually produce observable signatures visible by CMB and Gravitational Waves experiments.

However, the same issue has not been investigated in the cyclic Universe. Previous analyses of the spectrum of the gravitational wave assume a BD vacuum for perturbations produced in the \textit{ekpyrotic} phase, enforcing the solution to converge to that of a plane wave in flat space at sufficiently short distances~\cite{Boyle:2003km}. As argued in Ref.~\cite{Steinhardt:2002ih}, this choice mostly relies on the classical treatment of the dynamics of the evolving modulus field at arbitrarily small length scales. However, just like in inflation, the choice may no longer be trivial if the dynamics of the modulus field is influenced by new physics at some characteristic energy scale $M$. As a result, one may wonder whether in the cyclic Universe, the choice of the vacuum state is somehow affected by the same level of "arbitrariness" as inflation, or if additional constraints can be derived based on the strong interconnection between the different phases and/or from the need to maintain consistency through cycles. A different, yet related, question is whether, also in the cyclic Universe, this potential degree of arbitrariness impacts the predictions for the tensor spectrum or if they remain robust under the choice of the vacuum state.

Here, we take a first step forward in the discussion and, considering a non-BD vacuum state in the dark energy phase of the previous cycle, we argue that this initial state imposes a more fundamental constraint on the cyclic model, primarily due to the backreaction effects that a non-BD state could have on the background evolution of the modulus field. 

To prove this point, following Refs.~\cite{Aravind:2013lra,Ashoorioon:2013eia,Holman:2007na}, we allow the second Bogoliubov coefficient $\beta_k$ to be non-zero. Notice that, since the Bogoliubov coefficient is related to $D_2$ by 
\begin{equation}
\beta_k = 2D_2\sqrt{(k/\pi)},
\end{equation}
and parameterizes deviations from the BD vacuum. As the effective theory becomes invalid at energy scales beyond those of new physics, we ensure that no modes are excited past this point, therefore requiring $\beta_k \to 0$ for $k > Ma(\tau_c)$. In addition, we choose $\beta_k$ of the form
\begin{equation}
    \beta_k \sim \beta_0 e^{-\frac{k^2}{M^2 a\left(\tau_c\right)^2}},
    \label{eq:non-BD-vscuum}
\end{equation}
where the energy scale of the effective theory is $M\sim 10^{-4}M_{\rm{Pl}}$ and $\beta_0$ is a proportionality factor. The effects produced by the backreaction in the dark energy-dominated phase set stringent limits on how large the non-vanishing $\beta_k$ can be. In particular, requiring that the energy density of the non-BD quanta does not overwhelm the energy density associated with the modulus field -- hence spoiling the model -- implies
\begin{equation}
    \rho_{non-BD}\sim\frac{1}{a_{i}^4}\int{\frac{d^3k}{(2\pi)^3}}\left|\beta_0\right|^2 k\sim\left(\frac{a(\tau_c)}{a_{i}(\tau)}\right)^4\left|\beta_0\right|^2 M^4\ll M_{\rm Pl}^2 H^2,
    \label{eq:rho_NBD}
\end{equation}
where $a_{i}$ refers to the value of the scale factor when we set our initial vacuum conditions, and we have used $M_{\rm PL}^2H^2$ as the energy density associated with the background evolution. We take the cutoff time $\tau_c$, beyond which we cannot be certain of the modulus field dynamics to be the start of radiation domination regime in the previous cycle $\tau_{r(pc)}$.
As long as $(a(\tau_{r(pc)})/a(\tau))^4 \ll 1 $, Eq.~\eqref{eq:rho_NBD} sets an upper bound for the parameter $\beta_0$:
\begin{equation}
    \beta_0 < \frac{H M_{\rm Pl}}{M^2}\sim 10^{-53}
\end{equation}
where we used that, during the dark energy phase, the Hubble parameter is approximately constant $H\simeq H_0=100\,h \simeq 2.1 h \times 10^{-42}$ GeV. Notice that this bound is independent of the cutoff time $\tau_c$ when starting the calculation in the dark energy phase. This is because of the three expansion phases preceding this epoch and the large net expansion of the scale factor from cycle to cycle, ensuring that $(a(\tau_c)/a(\tau_i))^4 \ll 1$ for any $\tau_c$.

We emphasise once again that avoiding problems with backreaction during the dark energy-dominated phase leads to an extremely restrictive constraint on deviations away from the BD vacuum. To gain a rough idea of how restrictive our bound is, we can compare constraints on the same cutoff scale obtained in inflationary cosmology. In that case, as highlighted in Ref.~\cite{Holman:2007na}, the restriction reads $\beta_0 < 10^{-6}$; i.e., about 47 orders of magnitude weaker than our bound. The reason for such a large difference lies in the fact that, although inflation and dark energy share several common aspects in terms of background dynamics, these two phases span energy scales that differ by over 100 orders of magnitude in characteristic energy density. The energy scale of inflation is much higher, making it way more challenging for perturbations to overwhelm the dynamics and allowing larger freedom for deviations away from BD.

To further appreciate the strength of our constraint, it is also worth briefly discussing what happens when setting the vacuum state in a different phase of the model rather than in the dark energy-dominated one. In particular, we focus on the \textit{ekpyrotic} phase, which, as highlighted multiple times in this work, is the phase where initial conditions are typically fixed, and the calculation of the spectrum of tensor perturbations is initiated. In this phase, the Hubble parameter drastically grows, see Eq.~\eqref{Hratio}, while the scale factor remains approximately constant. As a consequence, the constraint on $\beta_0$ is relaxed up to $\beta_0 < 10^{45}$ for $M \sim 10^{-4}M_{\rm Pl}$, allowing us complete freedom to choose a wide range of vacuum states in the \textit{ekpyrotic} phase without compromising the model during this stage. That being said, it is necessary to ensure that significant deviations away from a BD vacuum during the \textit{ekpyrotic} phase do not lead to other issues during the subsequent evolutionary states of the cycle. However, given that the evolution of the dark energy phase is governed by the background dynamics of the lowest energy scale and considering the cyclic nature of the model and its resilience through cycles (meaning that starting from the dark energy phase of the previous cycle does not produce differences in observable quantities as we prove in the next subsection), we argue that selecting the vacuum during the dark energy phase is the most restrictive and conservative choice to circumvent this issue entirely. This makes relying on the BD vacuum a safer assumption from a model-building perspective.

\subsection{Gravitational waves produced from different phases}\label{analysis of gravitational waves}

Our general calculation for the strain spectrum of gravitational waves enables us to investigate how the predictions change when incorporating the contribution from tensor modes originating in the dark energy phase of the previous cycle. We can then test the robustness of these predictions by comparing our results with those already documented in the literature, derived from starting in the \textit{ekpyrotic} contracting phase.

In this section, we compare the predictions for the strain spectrum of gravitational waves obtained in the following two cases:
\begin{itemize}
\item[\textit{(a)}] Using our general calculation and starting in the dark energy phase of the previous cycle.  In this case we impose BD vacuum conditions on the coefficients of the dark energy stage $D_1$ and $D_2$ given by:
\begin{align}\label{DE conditions}
D_1 &= \frac{1}{2}\sqrt{\frac{\pi}{k}}, & D_2 &= 0.
\end{align}

\item[\textit{(b)}] Starting the calculation in the \textit{ekpyrotic} phase (disregarding the matching at $\tau=\tau_{tr}$) and imposing BD initial conditions as done in Res.~\cite{Boyle:2003km} 
\begin{align}\label{ekpyrotic conditions}
    A_1 &= \frac{1}{2}\sqrt{\frac{\pi}{k}}, & A_2 &= 0.
\end{align}
\end{itemize}
Notice that, although our calculation is fully generic concerning the choice of the vacuum state, as we proved in the previous subsection, deviations away from the BD vacuum state in the dark energy phase are strongly constrained, providing us with a valid physical reason to impose this vacuum state in the dark energy phase for the case \textit{(a)}. Instead, for the case \textit{(b)}, we impose the BD initial conditions in the \textit{ekpyrotic} phase to work in the same framework as Ref.~\cite{Boyle:2003km} and allow direct comparison.

\begin{figure}
    \centering
    \includegraphics[width=0.7\columnwidth]{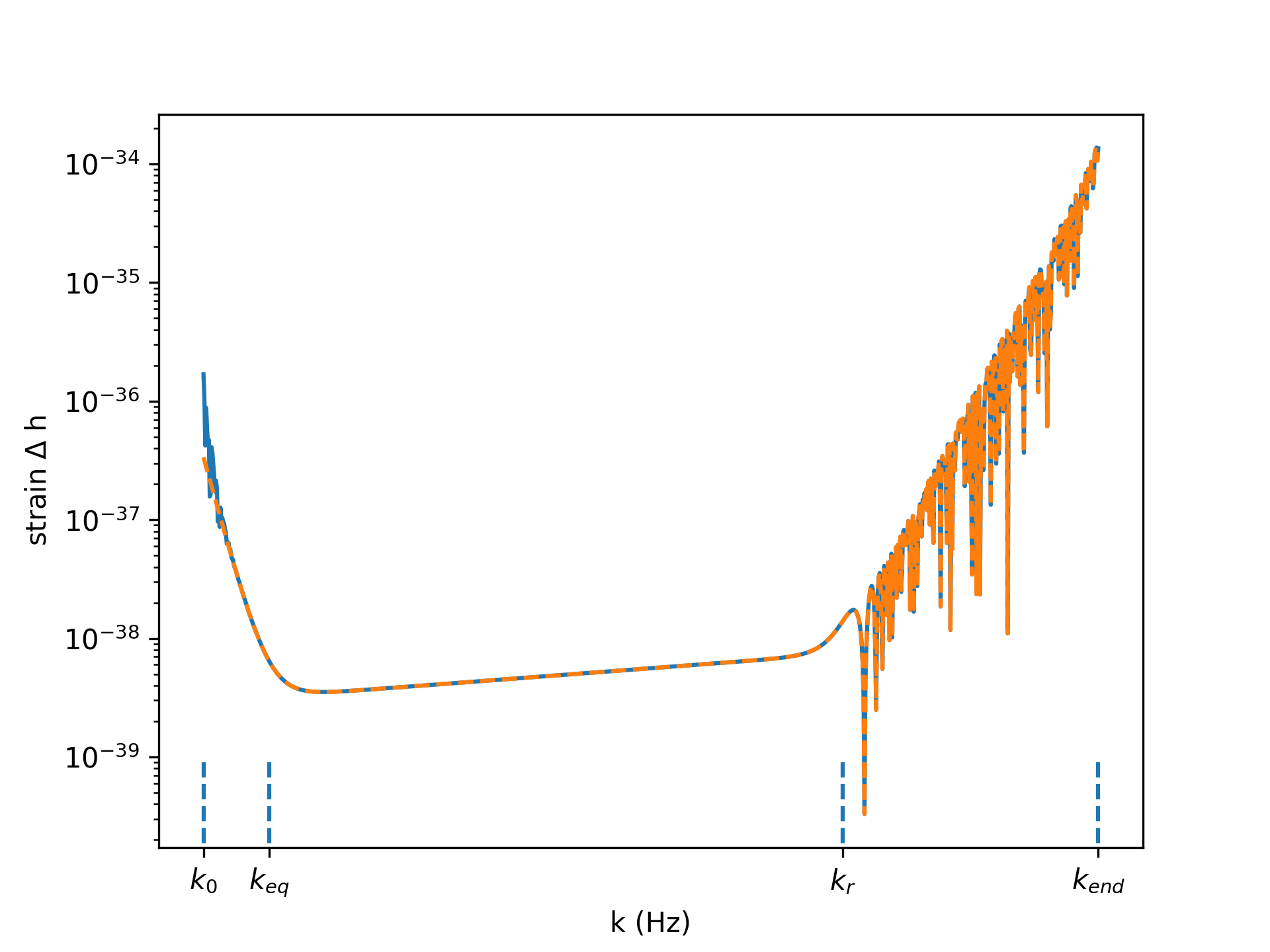}
    \caption{Strain spectrum plotted against comoving frequency $k$, of gravitational waves produced starting the cycle with the dark energy phase in blue, and with the \textit{ekpyrotic} phase in orange. BD initial conditions are assumed for both spectra. $k_0, k_{eq}, k_r$ and $k_{end}$ are the comoving frequencies on the horizon today, at matter-radiation equivalence, at the start of radiation domination, and at the \textit{ekpyrotic}-kinetic transition respectively.}
    \label{fig:Strain spectra}
\end{figure}

\begin{figure}[htb!]
    \centering
    \includegraphics[width=0.75\textwidth]{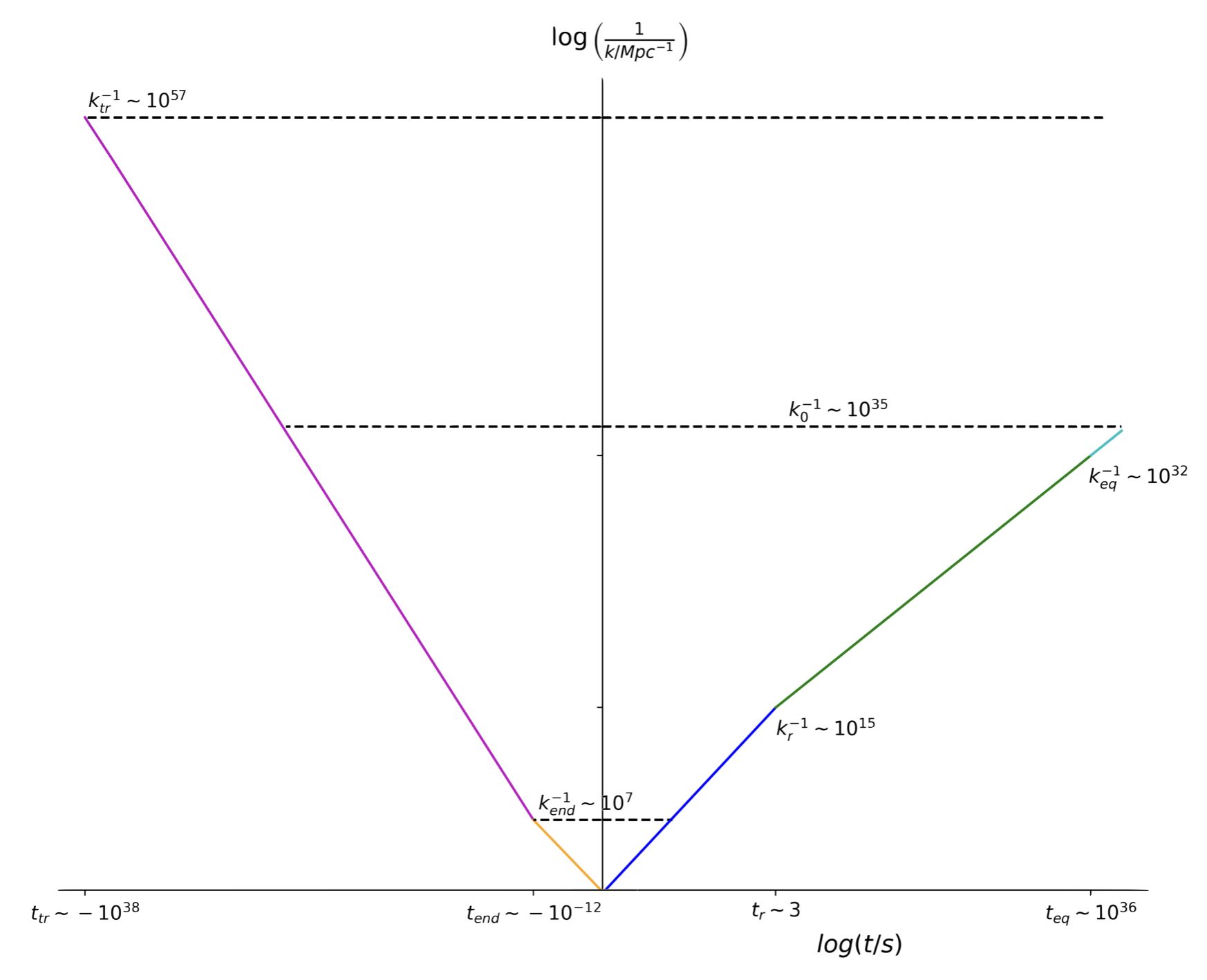}
    \caption{Illustrative plot of the comoving Hubble horizon, $1/aH$, throughout the \textit{ekpyrotic}, kinetic contracting, kinetic expanding, radiation domination, and finally matter domination phases in magenta, orange, dark blue, green and cyan respectively. Key modes on the horizon are illustrated as horizontal dashed lines, along with a label of their comoving wavenumber, and some less important modes are labelled.}
    \label{fig:horizon plot}
\end{figure}

After matching the relevant phases, the strain spectra of tensor modes predicted in the two cases can be derived using Eq.~\eqref{strain_spectrum} and Eq.~\eqref{transfer function} and are shown in \autoref{fig:Strain spectra} in blue for the case \textit{(a)} and in orange for the case \textit{(b)}. As evident from the figure, a difference of up to an order of magnitude in the strain $\Delta h$ is observed for modes $k_0$ at the present-day horizon.

This difference can be understood by considering the evolution of modes produced during the dark energy phase of the previous cycle. To further clarify this point, we refer to \autoref{fig:horizon plot} where we show the evolution of the comoving Hubble horizon, $1/aH$, throughout the different phases of the model. From the figure we note that the horizon at the end of the dark energy phase (\textit{i.e.} $k_{\rm tr}^{-1}$) is several orders of magnitude greater than the present-day horizon (\textit{i.e.} $k_0^{-1}$) ensuring that none of these super horizon modes can have re-entered by today or in any subsequent cycle.  Modes produced during this dark energy phase that exit the horizon become frozen and subsequently experience further stretching throughout this epoch.

On the other hand, sub-horizon modes produced in the same phase oscillate with decaying amplitude $h \propto a^{-1}$. This feature can be shown by solving Eq.~\eqref{BesselsEq} in the dark energy background dynamics described by Eq.~\eqref{adarkenergy}. In view of that, we expect sub-horizon modes produced deep within the dark energy phase (as well as in any previous phase of the previous cycle) to decay away to negligible amplitudes when compared to modes produced at the end of the same phase or during the subsequent \textit{ekpyrotic} phase. Instead, sub-horizon modes produced \textit{near} the end of the dark energy phase can survive into the \textit{ekpyrotic} phase of the next cycle, acting as extra quanta in the vacuum initial conditions for the subsequent \textit{ekpyrotic} phase. This contribution is encoded in the coefficient $A_2$ in Eq.~\eqref{ekpyrotic solution} and can lead to observable effects on scales corresponding to the long wavelength portion of the strain spectrum. Referring back to \autoref{fig:Strain spectra}, we can appreciate how the differences produced by amplitude oscillations expected from an under-damped simple harmonic oscillator solution affect only the range of frequencies between $k_0$ and $k_{\rm eq}$ before becoming frozen in during the \textit{ekpyrotic} phase.

Moving forward, as outlined in Ref.~\cite{Erickson:2006wc}, the subsequent phases will provide a red-tilt to the scale-invariant dark energy spectrum. Notice that for an \textit{ekpyrotic} phase lasting $N \sim 115$ e-folds, the magnitude of the coefficient $A_2$ becomes $\mathcal{O}(10^{10})$ between $k_0$ and $k_{\rm eq}$ then rapidly decays to order $\mathcal{O}(10^{-15})$ for modes around $k_{\rm r}$. This explains the discrepancy between imposing BD initial conditions in the dark energy phase given by Eq.~\eqref{DE conditions}, and in the \textit{ekpyrotic} phase, given by Eq.~\eqref{ekpyrotic conditions}. 

We conclude with a last important remark: as argued in \autoref{Gravitational Waves From A non-Bunch Davies Vacuum}, to preserve the background evolution of the modulus field, it is important to have a BD-like vacuum state during the dark energy phase. This requirement implies that perturbations existing in the current cycle must decay to negligible levels in amplitude compared to the energy density present in the BD vacuum, preventing them from acting as additional quanta on top of the vacuum. We emphasise that this condition is satisfied by the evolution of tensor perturbations after the bounce, particularly during the subsequent radiation and matter-dominated phases. In particular, the transfer function, Eq.~\eqref{transfer function}, ensures that the amplitudes of short-wavelength modes re-entering the horizon first (that are potentially the most problematic), are decreased by a factor $\left( k_0/k \right)^2$. Consequently, the amplitude of the shortest wavelength modes in the observable spectrum ($k \sim k_{\rm end}$), is suppressed by more than $20$ orders of magnitude during the radiation and matter-dominated phases. As a result, the evolution through these phases guarantees the decaying amplitude of all modes in every cycle preceding each dark energy phase and restoring the vacuum to a BD state in the latter. This shields the dark energy phase from possible backreaction effects, underscoring the resilience of the model and demonstrating once again that, from a theoretical standpoint, fixing initial conditions in this phase is much safer from a model-building perspective.

\section{Conclusion}
\label{sec:conc}

In this study, we investigate the production of gravitational waves in a cyclic Universe, focusing on certain conceptual aspects of the theory, such as the resilience of observable predictions against the phase of the cycle in which initial conditions are set and the choice of the vacuum state.

In most of the analyses carried out in the literature, the \textit{ekpyrotic} phase is typically assumed as the starting point of the cycle where initial conditions of primordial perturbations are imposed, and the calculations of the relative spectra begin.  While for scalar perturbations the implications of setting the initial conditions in different phases of the theory have been examined in a few studies surrounding this topic, to the best of our knowledge, the calculation of the tensor spectrum has always been performed starting in the \textit{ekpyrotic} phase, assuming a BD vacuum state and neglecting any potential contributions arising from tensor modes originating during the dark energy phase of the previous cycle.

In light of this, a few (we believe) interesting questions remained somewhat pending. For instance, one might wonder whether setting initial conditions in the dark energy phase of the previous cycle and considering the potential contribution of tensor modes generated in this phase could lead to any observable differences in the gravitational wave strain spectrum observed in the current cycle and eventually clarify why (not). Most importantly, in effective field theory descriptions of the cyclic Universe, the choice of the vacuum state may involve integrating out high-energy degrees of freedom, focusing on low-energy behaviors of the modulus field. This makes the choice far from trivial when considering the possibility of new physics acting at sufficiently high characteristic energy scales. Consequently, one might wonder about the implications of assuming a BD vacuum state in the \textit{ekpyrotic} phase and, more broadly, what freedom exists in the cyclic Universe regarding the choice of the vacuum state and how such freedom affects the predictions for the spectrum of gravitational waves.

Fuelled by these questions, we consider a cyclic Universe described by the effective 4D Lagrangian \eqref{lagrangian} with a potential given by Eq.~\eqref{potnetial}. After reviewing the background dynamics of the model (\autoref{construction of the model}), we focus on the production and evolution of tensor modes. In \autoref{primordial tensor spectrum}, we analytically solve the equation of motion of the gravitational wave field through all the different phases of the cycle by starting from the dark energy phase of the previous cycle and making no assumptions about the vacuum state. The results presented in this section generalise the predictions for the tensor spectrum to include the contribution of tensor perturbations produced during the dark energy phase of the previous cycle and apply to any choice of the vacuum state of the theory. Therefore, they extend the treatment presented so far in the literature, allowing us to gain important insights into the issues posed earlier.

As argued in \autoref{sec:discussion}, our findings reveal a significant resilience of the cyclic Universe model concerning predictions for the spectrum of primordial gravitational waves, with the most relevant results reading as follows.
\begin{itemize}
\item \textit{Initial Conditions:} To quantify the impact of the contribution arising from tensor modes produced in the previous dark energy phase of the cycle, in \autoref{analysis of gravitational waves} we compared the spectra obtained by starting from the \textit{ekpyrotic} contracting phase (in orange in \autoref{fig:Strain spectra}) and the dark energy phase of the previous cycle (in blue in \autoref{fig:Strain spectra}), assuming in both cases a BD vacuum. The two spectra are essentially identical except for a difference up to an order of magnitude in the strain $\Delta h$ for modes on the scale $k_0$ at the present-day horizon. The reason is that the horizon at the end of the dark energy phase is several orders of magnitude greater than the present-day horizon (see also \autoref{fig:horizon plot}), implying that none of these super-horizon modes can have re-entered by today or in any subsequent cycle. On the other hand, sub-horizon modes produced in the same phase oscillate with decaying amplitude $h \propto a^{-1}$. Therefore, sub-horizon modes produced deep within the dark energy phase (as well as in any previous phase of the previous cycle) decay away to negligible amplitudes, while sub-horizon modes produced near the end of the dark energy phase can survive into the \textit{ekpyrotic} phase of the next cycle, being responsible for the small differences in the strain spectrum observed around $k_0$.

\item \textit{Vacuum state:} Although including tensor modes originating from the previous dark energy phase does not lead to significant differences in the spectrum of primordial perturbations, we argue that starting the calculation in this phase imposes a more fundamental constraint on the choice of the vacuum state of the model. To highlight this aspect, in \autoref{Gravitational Waves From A non-Bunch Davies Vacuum} we consider a non-BD vacuum state in the dark energy phase parametrized in terms of the second Bogoliubov coefficient given by Eq.~\eqref{eq:non-BD-vscuum}. We show that requiring the energy density of the non-BD quanta not to overwhelm the energy density associated with the modulus field -- hence, potentially spoiling the model -- implies an extremely restrictive constraint on deviations away from the BD vacuum. On the other hand, for the same non-BD vacuum state, the constraint is completely lost in the \textit{ekpyrotic} phase where we are left with complete freedom to choose a wide range of vacuum states without compromising the model during this stage (but possibly spoiling the subsequent evolution). As a result, selecting the vacuum during the dark energy phase (where we are basically forced to BD) seems to be the most restrictive and conservative choice, as well as a safer assumption from a model-building perspective. Given the cyclic nature of the model and the resilience of its predictions against the phase where we set initial conditions (as proven in the previous point), this strongly reduces our freedom to consider exotic vacuum states in the cyclic Universe.
\end{itemize}

\begin{acknowledgments}
WG and CvdB are supported by the Lancaster–Sheffield Consortium for Fundamental Physics under STFC grant: ST/X000621/1. This article is based upon work from COST Action CA21136 Addressing observational tensions in cosmology with systematics and fundamental physics (CosmoVerse) supported by COST (European Cooperation in Science and Technology). 
\end{acknowledgments}

\bibliographystyle{JHEP}
\bibliography{bibliography}

\end{document}